\documentclass[12pt]{article}
\usepackage{graphicx}

\usepackage{epsf}
\def\beq{\begin{equation}}
\def\eeq{\end{equation}}
\def\bea{\begin{eqnarray}}
\def\eea{\end{eqnarray}}
\def\ve{\vert}
\def\vel{\left|}
\def\ver{\right|}
\def\nnb{\nonumber}
\def\ga{\left(}
\def\dr{\right)}

\def\rar{\rightarrow}
\def\nnb{\nonumber}
\def\la{\langle}
\def\ra{\rangle}
\def\ba{\begin{array}}
\def\ea{\end{array}}
\def\bea{\begin{eqnarray}}
\def\eea{\end{eqnarray}}

\def\ve{\vert}
\def\vel{\left|}
\def\ver{\right|}
\def\nnb{\nonumber}
\def\ga{\left(}
\def\dr{\right)}

\def\rar{\rightarrow}
\def\nnb{\nonumber}
\def\la{\langle}
\def\ra{\rangle}
\def\lla{\left<}
\def\rra{\right>}
\begin{document}
\title{ {\Large {\bf
New physics effects in the rare $B_s \rar \gamma \, \ell^+ \ell^-$ decays with polarized photon }}}
\author{ {\small U. O. Y{\i}lmaz$^1$ \,,
B. B.  \c{S}irvanl{\i}$^2$  \,\, and \,\,
G. Turan$^1$}\\
{\small $^1$ Physics Department, Middle East Technical University} \\
{\small 06531 Ankara, Turkey}\\
{\small $^2$ Gazi University, Faculty of Arts and Science, Department of Physics} \\
{\small 06100, Teknikokullar Ankara, Turkey}}

\begin{titlepage}
\maketitle
\thispagestyle{empty}
\begin{abstract}
Using the most general model independent form of the effective Hamiltonian,
the rare $B_s \rar \gamma \,\ell^+ \ell^- $ decays are studied  by taking into account the
polarization  of the photon. The  total and the differential branching ratios
for these decays, when photon is in the positive and negative helicity states, are presented.
Dependence of these observables   on the new Wilson coefficients are studied.
It is  also investigated the sensitivity of "photon polarization asymmetry" in
$B_s \rar \gamma \,\ell^+ \ell^- $
decays to the new Wilson coefficients. It has been shown that all these physical observables  are very sensitive to the
existence of new physics beyond SM and their experimental measurements can give valuable information
about it.

\noindent PACS number(s): 12.60.Fr, 13.20.He
\end{abstract}
\end{titlepage}
\section{Introduction}
The rare B-meson decays  induced by the flavor--changing neutral
currents (FCNC) have always been  important channels for obtaining
information about the fundamental parameters of the standard model
(SM), testing its predictions at loop level and probing possible
new physics.

The observation of radiative penguin mediated processes, in both
the exclusive $B \rar K^* \gamma$\cite{Ammar} and inclusive $B
\rar X_s \gamma$ \cite{CLEO} channels have prompted  the
investigation of the radiative rare B meson decays with a new
momentum. Among these, the semileptonic
$B_s \rar \gamma \, \ell^+ \ell^-$   $(\ell =e,\mu ,\tau)$ decays  have received a special interest due to their
relative cleanliness and sensitivity to new physics as well as ongoing experiments at the two B-factories
\cite{BFAC1,BFAC2}. It is well known that corresponding pure leptonic processes
$B_{s} \rar \ell^+ \ell^-$  have helicity suppression so that their
decay width  are too small to be measured for the light lepton modes. In SM the branching ratio of the
$BR(B_s \rar e^+ e^-,~\mu^+ \mu^-)\simeq 4.2 \times 10^{-14}$ and
$1.8 \times 10^{-9}$, respectively. Although $\tau$ channel is free from this
suppression, its experimental detection is quite hard due to the low
efficiency. In $B_s \rar \tau^+ \tau^-\gamma $ decay,
 helicity suppression is overcome by the photon emission in addition to the
lepton pair. Therefore,  it is expected for $B_s \rar \gamma \ell^+ \ell^- $ decay  to have
a larger branching ratio and this makes its investigation  interesting. Indeed, $B_s \rar \gamma \ell^+ \ell^- $
decays
have been widely investigated in the framework of the SM for light and heavy
lepton modes \cite{Eilam1}-\cite{Geng}, and reported
$BR(B_s \rar \gamma \, e^+ e^-,~\gamma \,\mu^+ \mu^-,~\gamma \,\tau^+ \tau^-)= 2.35 \times 10^{-9}$ ,
$1.9 \times 10^{-9}$ and $9.54 \times 10^{-9}$, respectively. The new physics
effects
in these decays have also been studied in some models, like MSSM \cite{Xiong}-\cite{Choud3} and the two Higgs doublet
model \cite{Iltan1}-\cite{Erkol2}, and shown that  different observables, like branching ratio, forward-backward asymmetry,
etc., are very sensitive to the physics beyond the SM.
Investigation of the polarization effects may provide another efficient way in establishing the new physics.
Along this line, the polarization asymmetries of the final state lepton in $B_s \rar \gamma \ell^+ \ell^- $
decays have been studied in MSSM in \cite{Choud1} and concluded that they can be very useful
for accurate determination of various Wilson coefficients.

In a radiative decay mode like ours, the final state photon can also
emerge with a definite polarization  and  provide another kinematical variable to study the new
physics effects \cite{Choud2}. In this paper, we will   study the rare $B_s \rar \gamma \,\ell^+
\ell^- $ decay  by taking into account the photon polarization. Although experimental measurement
of this variable would be much more difficult than that of e.g.,
polarization  of the final leptons in $B_s \rar \gamma \, \ell^+ \ell^- $ decay,
this is still another kinematical variable for
studying radiative decays. In our work we will investigate
sensitivity of such "photon polarization asymmetry" in $B_s \rar
\gamma \,\ell^+ \ell^- $ decay to the new Wilson coefficients, in addition to
the study of total and differential branching ratios with polarized final state photon.
Doing this we use a most general model independent  effective Hamiltonian, which contains
the scalar and tensor type interactions as well as the vector types (See Eq.(\ref{effH}) below).
We note that in a recent work \cite{Berin} we have studied  a related mode $B_s \rar \gamma \,\nu \bar{\nu}$
with a polarized photon in a similar way  and showed  that the spectrum is sensitive to the types of the
interactions so that it is useful to discriminate the various new physics effects.

The paper is organized as follows: In section 2, we present the most general, model
independent form of the effective Hamiltonian and the parametrization of the
hadronic matrix elements in terms of appropriate form factors. We then calculate
the differential decay width and the  photon polarization  asymmetry for the
$B \rar \gamma \, \ell^+ \ell^- $ decay when the photon is in
positive and negative helicity states . Section 3 is devoted to the numerical analysis and
discussion of our results.

\section{Matrix element for the  $B_s  \rar \gamma \, \ell^+ \ell^- $ decay \label{s2}}
The matrix element for the process $B_s \rar \gamma \, \ell^+\ell^-$ can be obtained from
that of the purely leptonic $B \rar \ell^+ \ell^-$ decay.  Therefore, we start with the
effective Hamiltonian for $b \rar  s  \ell^+ \ell^- $ transition written in terms of
twelve model independent four-Fermi interactions as follows \cite{Aliev3b}:
\bea
\label{effH}
{\cal H}_{eff} &=& \frac{G\alpha}{\sqrt{2} \pi}
 V_{ts}V_{tb}^\ast
\Bigg\{ C_{SL} \, \bar s i \sigma_{\mu\nu} \frac{q^\nu}{q^2}\, L \,b
\, \bar \ell \gamma^\mu \ell + C_{BR}\, \bar s i \sigma_{\mu\nu}
\frac{q^\nu}{q^2} \,R\, b \, \bar \ell \gamma^\mu \ell \nnb \\
&&+C_{LL}^{tot}\, \bar s_L \gamma_\mu b_L \,\bar \ell_L \gamma^\mu \ell_L +
C_{LR}^{tot} \,\bar s_L \gamma_\mu b_L \, \bar \ell_R \gamma^\mu \ell_R +
C_{RL} \,\bar s_R \gamma_\mu b_R \,\bar \ell_L \gamma^\mu \ell_L \nnb \\
&&+C_{RR} \,\bar s_R \gamma_\mu b_R \, \bar \ell_R \gamma^\mu \ell_R +
C_{LRLR} \, \bar s_L b_R \,\bar \ell_L \ell_R +
C_{RLLR} \,\bar s_R b_L \,\bar \ell_L \ell_R \\
&&+C_{LRRL} \,\bar s_L b_R \,\bar \ell_R \ell_L +
C_{RLRL} \,\bar s_R b_L \,\bar \ell_R \ell_L+
C_T\, \bar s \sigma_{\mu\nu} b \,\bar \ell \sigma^{\mu\nu}\ell \nnb \\
&&+i C_{TE}\,\epsilon^{\mu\nu\alpha\beta} \bar s \sigma_{\mu\nu} b \,
\bar \ell \sigma_{\alpha\beta} \ell  \Bigg\}~, \nnb
\eea
where  $L$ and $R$  are the chiral projection operators defined as $(1\pm \gamma_5 )/2$,
respectively. In (\ref{effH}),
$C_X$ are the coefficients of the four--Fermi interactions with $X=LL,LR,RL,RR$
describing vector, $X=LRLR,RLLR,LRRL,RLRL$ scalar and $X=T,TE$ tensor type interactions.
We note that several of the Wilson coefficients in Eq. (\ref{effH}) do already exist in the SM:
in the SM, $C_{LL}$ and $C_{LR}$ are in the form
$C_9^{eff} - C_{10}$ and $C_9^{eff} + C_{10}$ for the $b\rar s \ell^+\ell^-$ decay, while
the coefficients $C_{SL}$ and $C_{BR}$ correspond to $-2 m_s C_7^{eff}$ and $-2
m_b C_7^{eff}$, respectively. Therefore, writing
\bea
C_{LL}^{tot} &=& C_9^{eff}-C_{10} + C_{LL}~, \nnb \\
C_{LR}^{tot} &=& C_9^{eff}+C_{10} + C_{LR}~, \nnb
\eea
we see that $C_{LL}^{tot}$ and $C_{LR}^{tot}$ contain the contributions from the SM and also from the new
physics.

Having established the general form of the effective Hamiltonian,
we proceed to calculate
the matrix element of the $B_s \rar \gamma \, \ell^+\ell^-$ decay.
This exclusive decay can receive short-distance contributions from
the box, Z, and photon penguin diagrams for $b \rar  s  $ transition  by attaching
an additional photon line to any internal or external lines. As
pointed out before \cite{Aliev1,Aliev2}, contributions coming from
the release of the free photon from  any charged internal line
will be suppressed by a factor of $m^2_b/M^2_W$ and we  neglect
them in the following analysis. When a photon is released from the
initial quark lines it contributes to the so-called "structure
dependent" (SD) part of the amplitude, ${\cal M}_{SD}$. Then, it
follows from Eq. (\ref{effH}) that, in order to calculate  ${\cal M}_{SD}$,
the  matrix elements  needed and their definitions in
term of the various form factors are as follows \cite{Aliev1,Eilam2}:
\bea
 \label{mel1}
 \lla \gamma(k) \vel
\bar s \gamma_\mu (1 \mp \gamma_5) b \ver B(p_B) \rra &=&
\frac{e}{m_B^2} \Big\{ \epsilon_{\mu\nu\lambda\sigma}
\varepsilon^{\ast\nu} q^\lambda
k^\sigma g(q^2) \nnb \\
&&\pm i\,
\Big[ \varepsilon^{\ast\mu} (k q) -
(\varepsilon^\ast q) k^\mu \Big] f(q^2) \Big\}~,\\ \nnb \\
\label{mel2}
\lla \gamma(k) \vel \bar s \sigma_{\mu\nu} b \ver B(p_B) \rra &=&
\frac{e}{m_B^2}
\epsilon_{\mu\nu\lambda\sigma} \Big[
G \varepsilon^{\ast\lambda} k^\sigma +
H \varepsilon^{\ast\lambda} q^\sigma +
N (\varepsilon^\ast q) q^\lambda k^\sigma \Big]~,
\eea
\bea
\label{mel1q}
\lla \gamma(k) \vel \bar s (1 \mp \gamma_5) b \ver B(p_B) \rra
&=& 0~, \\
\label{mel2q}
\lla \gamma \vel \bar s i \sigma_{\mu\nu} q^\nu b \ver B (p_B) \rra &=&
\frac{e}{m_B^2} i\, \epsilon_{\mu\nu\alpha\beta} q^\nu
\varepsilon^{\alpha\ast} k^\beta G~,
\eea
and
\bea
\label{mel22q}
\lla \gamma (k) \vel \bar s i \sigma_{\mu\nu} q^\nu (1+\gamma_5) b \ver B (p_B)\rra &=&
\frac{e}{m_B^2} \Big\{
\epsilon_{\mu\alpha\beta\sigma} \, \varepsilon^{\alpha\ast} q^\beta k^\sigma
g_1(q^2)
+ i\,\Big[\varepsilon_\mu^\ast (q k) - (\varepsilon^\ast q) k_\mu \Big]
f_1(q^2) \Big\}~, \nnb \\ & &
\eea
where $\varepsilon_\mu^\ast$ and $k_\mu$ are the four vector polarization
and four momentum of the photon, respectively, $q$ is the momentum transfer,
 $p_B$ is the momentum of the $B$ meson, and $G,~H$ and $N$  can be
expressed  in terms of the form factors $g_1$ and $f_1$ by using Eqs. (\ref{mel2}),
(\ref{mel2q}) and (\ref{mel22q}). The
matrix element  describing the structure--dependent part can be obtained
from Eqs. (\ref{mel1})--(\ref{mel22q}) as
\bea
\label{sd}
{\cal M}_{SD} &=& \frac{\alpha G_F}{4 \sqrt{2} \, \pi} V_{tb} V_{ts}^*
\frac{e}{m_B^2} \,\Bigg\{
\bar \ell \gamma^\mu (1-\gamma_5) \ell \, \Big[
A_1 \epsilon_{\mu \nu \alpha \beta}
\varepsilon^{\ast\nu} q^\alpha k^\beta +
i \, A_2 \Big( \varepsilon_\mu^\ast (k q) -
(\varepsilon^\ast q ) k_\mu \Big) \Big] \nnb \\
&+& \bar \ell \gamma^\mu (1+\gamma_5) \ell \, \Big[
B_1 \epsilon_{\mu \nu \alpha \beta}
\varepsilon^{\ast\nu} q^\alpha k^\beta
+ i \, B_2 \Big( \varepsilon_\mu^\ast (k q) -
(\varepsilon^\ast q ) k_\mu \Big) \Big] \nnb \\
&+& i \, \epsilon_{\mu \nu \alpha \beta}
\bar \ell \sigma^{\mu\nu}\ell \, \Big[ G \varepsilon^{\ast\alpha} k^\beta
+ H \varepsilon^{\ast\alpha} q^\beta +
N (\varepsilon^\ast q) q^\alpha k^\beta \Big] \\
&+& i \,\bar \ell \sigma_{\mu\nu}\ell \, \Big[
G_1 (\varepsilon^{\ast\mu} k^\nu - \varepsilon^{\ast\nu} k^\mu) +
H_1 (\varepsilon^{\ast\mu} q^\nu - \varepsilon^{\ast\nu} q^\mu) +
N_1 (\varepsilon^\ast q) (q^\mu k^\nu - q^\nu k^\mu) \Big] \Bigg\}~,\nnb
\eea
where
\bea
A_1 &=& \frac{1}{q^2} \Big( C_{BR} + C_{SL} \Big) g_1 +
\Big( C_{LL}^{tot} + C_{RL} \Big) g ~, \nnb \\
A_2 &=& \frac{1}{q^2} \Big( C_{BR} - C_{SL} \Big) f_1 +
\Big( C_{LL}^{tot} - C_{RL} \Big) f ~, \nnb \\
B_1 &=& \frac{1}{q^2} \Big( C_{BR} + C_{SL} \Big) g_1 +
\Big( C_{LR}^{tot} + C_{RR} \Big) g ~, \nnb \\
B_2 &=& \frac{1}{q^2} \Big( C_{BR} - C_{SL} \Big) f_1 +
\Big( C_{LR}^{tot} - C_{RR} \Big) f ~, \nnb \\
G &=& 4 C_T g_1 ~~~,~~~ N = - 4 C_T \frac{1}{q^2} (f_1+g_1) ~, \nnb \\
H &=& N (qk) ~~~,~~~G_1 = - 8 C_{TE} g_1 ~, \nnb \\
N_1 &=& 8 C_{TE} \frac{1}{q^2} (f_1+g_1) ~~~,~~~ H_1 = N_1(qk)~ .\nnb
\eea

When photon is radiated from the lepton line we get the the so-called "internal  Bremsstrahlung" (IB) contribution,
${\cal M}_{IB}$. Using the expressions
\bea
\la 0 \ve \bar s \gamma_\mu \gamma_5 b \ve B (p_B)\ra &=&
-~i f_B p_{B\mu}~, \nnb \\
\la 0 \ve \bar s \sigma_{\mu\nu} (1+\gamma_5) b \ve B (p_B) \ra &=& 0~,\nnb
\eea
and conservation of the vector current, we get
\bea
\label{ib}
{\cal M}_{IB} &=& \frac{\alpha G_F}{4 \sqrt{2} \, \pi} V_{tb} V_{ts}^*
e f_B i \,\Bigg\{
F\, \bar \ell  \Bigg(
\frac{{\not\!\varepsilon}^\ast {\not\!p}_B}{2 p_1 k} -
\frac{{\not\!p}_B {\not\!\varepsilon}^\ast}{2 p_2 k} \Bigg)
\gamma_5 \ell \nnb \\
&+& F_1 \, \bar \ell  \Bigg[
\frac{{\not\!\varepsilon}^\ast {\not\!p}_B}{2 p_1 k} -
\frac{{\not\!p}_B {\not\!\varepsilon}^\ast}{2 p_2 k} +
2 m_\ell \Bigg(\frac{1}{2 p_1 k} + \frac{1}{2 p_2 k}\Bigg)
{\not\!\varepsilon}^\ast \Bigg] \ell \Bigg\}~,
\eea
where
\bea
F &=& 2 m_\ell \Big( C_{LR}^{tot} - C_{LL}^{tot} + C_{RL} - C_{RR} \Big)
+ \frac{m_B^2}{m_b}
\Big( C_{LRLR} - C_{RLLR} - C_{LRRL} + C_{RLRL} \Big)~, \nnb \\
F_1 &=&\frac{m_B^2}{m_b} \Big( C_{LRLR} - C_{RLLR} + C_{LRRL} - C_{RLRL}
\Big)~.
\eea
Finally, the total matrix element for the $B_s \rar \gamma \, \ell^{+}\ell^{-}$ decay
is obtained as a sum of the ${\cal M}_{SD}$ and ${\cal M}_{IB}$ terms,
\beq
{\cal M}={\cal M}_{SD}+{\cal M}_{IB}.
\eeq

The next task is the calculation of the  differential decay rate
of $B_s \rar \gamma \, \ell^+ \ell^- $ decay  as a function of
dimensionless parameter $x=2 E_{\gamma}/m_B$, where $E_{\gamma}$
is the  photon energy. In the center of mass (c.m.) frame of the
dileptons $\ell^+\ell^-$, where we take  $z=\cos \theta $ and $\theta$ is the angle
between the momentum of the $B_{s}$-meson and that of $\ell^-$, double differential decay width
is found to be
\bea
\label{dGdxdz}
\frac{d \Gamma}{dx \, dz} = \frac{1}{(2 \pi)^3 64 }\, x \, v \, m_B \, \vel {\cal M} \ver^2~,
\eea
with
\bea
\vel {\cal M} \ver^2 & = &\vel {\cal M}_{SD} \ver^2+\vel {\cal M}_{IB} \ver^2+
2 Re({\cal M}_{SD}{\cal M}^{*}_{IB} )\label{M2}
\eea
where $v=\sqrt{1-\frac{4 r}{1-x}}$ and $r=m^2_{\ell}/m^2_{B}$.
We note that $\vel {\cal M}_{IB} \ver^2 $ term has infrared singularity due to
the emission of soft photon. In order to obtain a finite result, we follow the approach
described in \cite{Aliev2} and impose a cut on the photon energy, i.e., we require
$E_{\gamma}\geq 25$ MeV, which corresponds to detect only hard photons experimentally.
This cut requires that $E_{\gamma}\geq \delta \, m_B /2$ with $\delta =0.01$.

In such a   radiative decay, the final
state photon can emerge with a definite polarization and there
follows the question of how sensitive the branching ratio is to
the new Wilson coefficients when the photon is in the positive or
negative helicity states.  To find an answer to this question, we
evaluate $\frac{d\Gamma (\varepsilon^\ast=\varepsilon_1)}{dx}$ and
$\frac{d\Gamma (\varepsilon^\ast=\varepsilon_2)}{dx}$ for $B_s
\rar \gamma \, \ell^+ \ell^- $  decay, in  the c.m. frame of
$\ell^+ \ell^- $, in which four-momenta and polarization vectors
, $\varepsilon_1$ and $\varepsilon_2$, are as follows:
\bea P_B & =
& (E_B,0,0,E_k) \,\,\, , \,\,\, k = (E_k,0,0,E_k) \,\,\, , \,\,\,
p_1=(p,0,p \sqrt{1-z^2},-p z) ~ ,\nnb \\ p_2 & = & (p,0,-p
\sqrt{1-z^2},p z)\,\,\, , \,\,\, \varepsilon_1 =
(0,1,i,0)/\sqrt{2}\,\,\, , \,\,\,\varepsilon_2 =
(0,1,-i,0)/\sqrt{2}~ , \label{mom}
\eea
where $E_B=m_B (2-x)/2
\sqrt{1-x}$, $E_k=m_B x/2 \sqrt{1-x}$, and $p=m_B \sqrt{1-x}/2$. Using the above forms,
we obtain
\bea
\label{bela}
\frac{d\Gamma (\varepsilon^\ast=\varepsilon_i)}{dx}&
= & \,\vel \frac{\alpha G_F}{4 \sqrt{2} \, \pi} V_{tb} V_{ts}^*
\ver^2 \, \frac{\alpha}{\ga 2 \, \pi \dr^3}\,\frac{\pi}{4}\,m_B \,
\Delta(\varepsilon_i)
\eea
where
\bea
\label{Deltapm}
\lefteqn{\Delta (\varepsilon _{i})
= \frac{v x}{3}\Bigg\{ 4 \,x \Big ( (8r+x) \vel H_1 \ver^2
-(4r-x) \vel H \ver^2 \Big )-6m_{\ell} (1-x)^2
\mbox{\rm Im} [(A_2 \pm A_1+B_2\pm B_1) G_1^\ast]} \nnb \\
&&+ \frac{2}{x} (1-x)^2 (2r+x)\Big (\vel G_1 \ver^2
+\vel G \ver^2 \pm 2 \mbox{\rm Im} [-G_1 G^\ast] \Big ) )-12
m_{\ell} (1-x)x
\mbox{\rm Im} [(A_2 \pm A_1+B_2\pm B_1) H_1^\ast] \nnb \\
& & \pm 4 (1-x)\Big ( (8r+x) \mbox{\rm Im} [G
H_1^\ast] +(4r-x) \mbox{\rm Im} [G_1 H^\ast]  \Big
)+6m^2_{\ell}(1-x)^2 \mbox{\rm Re} [(A_1\pm A_2)(B_1\pm B_2)] \nnb
\\ & & +
 m^2_B (1-x)^2 (x-r) \Big (  \vel A_1
\ver^2+ \vel A_2 \ver^2+ \vel B_1 \ver^2+ \vel B_2 \ver^2 \pm 2
\mbox{\rm Re} [A_1 A_2^\ast +B_1 B_2^\ast]\Big ) \nnb \\ & &-
6 m_{\ell}(1-x)^2 \mbox{\rm Re} [(A_2
\pm A_1+B_2\pm B_1) G^\ast]+4 (1-x) \Big ( (8r+x) \mbox{\rm
Re}[G_1 H_1^\ast]-(4r-x) \mbox{\rm Re}[G H^\ast]\Big ) \Bigg\}
\nnb \\ & & +
 \frac{2 x}{(1-x)^2}f_B^2 \Bigg\{(-2 v x+(1-4 r+x^2) \mbox{\rm ln}[u]) \vel F \ver^2 \pm 2 (1-x) (2 v x-(1-4 r+x) \mbox{\rm ln}[u]) \mbox{\rm Re}[F F_1^\ast] \nnb \\
 & & +  \Big ( 2 v x (4 r
-1)+(1+16 r^2+x^2-4 r (1+2 x)) \mbox{\rm ln}[u] \vel F_1 \ver^2
\Big )\Bigg\} \nnb \\  & & +  2 x
f_B \Bigg\{\pm (v x+2 r \mbox{\rm ln}[u] ) \mbox{\rm Im}[-F
H_1^\ast] \pm m_{\ell} (1-x) \mbox{\rm ln}[u] \mbox{\rm Re}[(A_2 \pm A_1+B_2\pm B_1) F^\ast] \nnb \\
& & -  m_{\ell} (2 v x+(1-4 r-x) \mbox{\rm ln}[u])
\mbox{\rm Re}[(A_2 \pm A_1+B_2\pm B_1) F_1^\ast]\nnb \\
& & - 2 (v-2
r \mbox{\rm ln}[u]) \mbox{\rm Im}[(-F_1\pm F)(G_1^\ast \pm
G^\ast)]\pm 2 (v x-2 r \mbox{\rm ln}[u])
\mbox{\rm Re}[F_1 H^\ast] \nnb \\
& & +
\frac{2}{(1-x)}\Big (  (v x (1+x)+2 r (1-3 x) \mbox{\rm ln}[u])
\mbox{\rm Im}[F_1 H_1^\ast]-(1+x) (v x-2 r \mbox{\rm ln}[u])
\mbox{\rm Re}[F_1 H^\ast]\Big ) \Bigg\}
\eea
where  $+(-)$ is for $i=1(2)$ and $u=1+v/1-v$.

The effects of polarized photon can be also studied through a variable "photon polarization asymmetry"
\cite{Choud2}:
\bea
H(x)=\frac{\frac{d\Gamma (\varepsilon^\ast=\varepsilon_1)}{dx}-
\frac{d\Gamma (\varepsilon^\ast=\varepsilon_2)}{dx}}{\frac{d\Gamma (\varepsilon^\ast=\varepsilon_1)}{dx}+
\frac{d\Gamma (\varepsilon^\ast=\varepsilon_2)}{dx}}
& = & \frac{\Delta(\varepsilon_1)-\Delta(\varepsilon_2)}{\Delta_0} \, ,
\eea
where
\bea
\label{Hx}
\lefteqn{\Delta(\varepsilon_1)-\Delta(\varepsilon_2) =
\frac{4}{3}  x^2 v  \Bigg\{\frac{2 x (1+2 r-x)}{(-1+x)} \mbox{\rm Im}
[G_1 G^\ast]-3 m_{\ell} x \Big ( \mbox{\rm Im} [(A_1+B_1)G_1^\ast]
+\mbox{\rm Re} [(A_2+B_2)G^\ast ] \Big ) }\nnb \\
& - &
6 m_{\ell} (1-x)( (\mbox{\rm Im}[(A_1+B_1) H_1^\ast])+2 \Big ( (1+8 r-x)
\mbox{\rm Im} [G H_1^\ast]-(1-4 r-x) \mbox{\rm Im} [G_1 H^\ast]\Big )
\nnb \\ & + & m_B^2 x \Big ( 3 r (\mbox{\rm Re} [A_2 B_1^\ast +A_1 B_2^\ast]
 +(1-r-x) \mbox{\rm Re} [B_1 B_2^\ast +A_1 A_2^\ast] \Big ) \Bigg\}
\nnb \\ & + & 8 f_B^2 \Big (  2 v (1-x)-(2-4 r- x) \mbox{\rm ln}[u]  \Big )+
4 f_B x \Bigg\{2 (v (x-1)-2 r \mbox{\rm ln}[u]) \mbox{\rm Im} [F H_1^\ast]
\nnb \\ & + & m_{\ell} x \mbox{\rm ln}[u] \mbox{\rm Re}[(A_2+B_2) F^\ast]+
m_{\ell} \Big (  2 v (x-1)+( 4 r- x) \mbox{\rm ln}[u]  \Big )
\mbox{\rm Re}[(A_1+B_1) F_1^\ast]\nnb \\ & + &
2 (v -2 r \mbox{\rm ln}[u]) \mbox{\rm Re}[F_1 G^\ast]- \mbox{\rm Im}
[F G_1^\ast]+2 ( v (1-x)-2 r \mbox{\rm ln}[u] ) \mbox{\rm Re}[F_1 H^\ast]\Bigg\}
\eea
and

\bea
\label{bela2}
\lefteqn{
\Delta_0 = \Bigg\{ x^3 v\, \Bigg (4 m_\ell \,
\mbox{\rm Re}\Big( [A_1+B_1] G^\ast\Big)
- \, 4 m_B^2 r \, \mbox{\rm Re}\Big( A_1 B_1^\ast + A_2 B_2^\ast \Big)} \nnb \\
&&-  4 \Big[ \vel H_1 \ver^2 (1-x) + \mbox{\rm Re}\Big( G_1 H_1^\ast \Big) x
\Big] \frac{(1+ 8 r - x)}{x^2} -  4  \Big[ \vel H \ver^2 (1-x) + \mbox{\rm Re}\Big( G H^\ast \Big) x
\Big] \frac{(1- 4 r - x)}{x^2} \nnb \\
&&+ \, \frac{1}{3} m_B^2 \Big[ 2\,\mbox{\rm Re}\Big( G N^\ast \Big) +
m_B^2 \vel N \ver^2 (1-x) \Big] (1-4r-x) \nnb \\
&&+ \, \frac{1}{3} m_B^2 \Big[ 2\,\mbox{\rm Re}\Big( G_1 N_1^\ast \Big) +
m_B^2 \vel N_1 \ver^2 (1-x) \Big] (1+8r-x) \nnb \\
&&- \, \frac{2}{3} m_B^2 \Big( \vel A_1 \ver^2 + \vel A_2 \ver^2 +
\vel B_1 \ver^2 + \vel B_2 \ver^2\Big)
( 1 - r - x) - \frac{4}{3} \Big( \vel G \ver^2 + \vel G_1 \ver^2 \Big)
\frac{(1 + 2 r - x)}{(1 - x)}\nnb \\
&&+\,2 m_\ell \,\mbox{\rm Im}\Big([A_2+B_2] [6 H_1^\ast (1-x)
 + 2 G_1^\ast x - m_B^2 \,N_1^\ast  x (1-x)]\Big)\frac{1}{x} \Bigg ) \nnb \\
&&+ \, 4 f_B \,\Bigg ( 2 v \,\Bigg[
\mbox{\rm Re}\Big(F G^\ast\Big)\frac{1}{(1-x)}
- \mbox{\rm Re}\Big(F H^\ast\Big)
+ \,m_B^2\, \mbox{\rm Re}\Big(F N^\ast\Big)
+ m_\ell \,\mbox{\rm Re}\Big([A_2+B_2] F_1^\ast\Big)
\Bigg] \, x (1-x)\nnb \\
&&+\, {\rm ln} [u] \Bigg[
m_\ell \, \mbox{\rm Re}\Big([A_2+B_2] F_1^\ast\Big) \, x (x-4 r)
+ 2  \, \mbox{\rm Re}\Big(F H^\ast \Big)
\Big[1 -x + 2 r (x-2) \Big] \nnb \\
&&- \, 4 r x\, \mbox{\rm Re}\Big(F G^\ast\Big)
+ m_B^2 \, \mbox{\rm Re}\Big(F N^\ast\Big) \, x (x-1)
- m_\ell \,\mbox{\rm Re}\Big([A_1+B_1] F^\ast\Big) \, x^2
\Bigg]\Bigg ) \nnb \\
&&+ \,  4 f_B^2
\Bigg (2 v\, \Big( \vel F \ver^2+
(1-4 r) \vel F_1 \ver^2 \Big) \frac{(1-x)}{x} \nnb \\
&&+ \, {\rm ln}[u] \Bigg[ \vel F \ver^2 \Big( 2 +\frac{4 r}{x} -\frac{2}{x} -x \Big)
+ \vel F_1 \ver^2 \Bigg( 2 (1-4 r) - \frac{2 \ga 1- 6 r + 8 r^2 \dr}{x} -x
\Bigg) \Bigg]\Bigg )\Bigg\}~.
\eea
The expression in Eq.(\ref{Hx}) agrees with \cite{Choud2} for the SM case with neutral Higgs contributions.
\section{Numerical analysis and discussion \label{s3}}

We present here our numerical analysis about the branching ratios (BR) and the photon polarization
asymmetries ($H$) for the $B_s \rar \gamma \ell^+ \ell^- $ decays with $\ell =\mu , \tau $.
We first give the input parameters used in our numerical analysis :
\begin{eqnarray}
& & m_B =5.28 \, GeV \, , \, m_b =4.8 \, GeV \, , \,m_{\mu} =0.105 \, GeV \, , \,
m_{\tau} =1.78 \, GeV \, , \nnb \\
& & f_B=0.2 \, GeV \, , \, \, |V_{tb} V^*_{ts}|=0.045 \, \, , \, \, \alpha^{-1}=137  \, \,  ,
G_F=1.17 \times 10^{-5}\, GeV^{-2} \nnb \\
& &  \tau_{B_{s}}=1.54 \times 10^{-12} \, s \, \, , \, \, C_9^{eff}=4.344 \, \, , \, \, C_{10}=-4.669  .
\end{eqnarray}
Furthermore we assume in this work that
all new Wilson coefficients are real and vary in the region $-4\leq C_X\leq 4$.
We note that such a  choice for the range of the new Wilson coefficients follows from
the experimental bounds on the branching ratios of the $B \rar K^{\ast}\mu^+\mu^-$ \cite{ABE} and
$B_s\rar \mu^+\mu^-$ decays \cite{Halyo}.
It should be  noted here  that the value of the Wilson coefficient $C^{eff}_9$ above corresponds
only to the short-distance contributions. $C^{eff}_9$ also receives long-distance
  contributions associated with the real $\bar{c}c$ intermediate states; but in this work
we consider  only  the short distance effects.

To make some numerical predictions, we also need the explicit forms of the form factors $g,~f,~g_1$ and $f_1$.
They are calculated in framework of light--cone $QCD$ sum rules
in \cite{Eilam2,Aliev1}, and also in \cite{Kruger} in terms of two parameters $F(0)$ and $m_F$.
In our work we have used the results of  \cite{Aliev1}, in which   $q^2$ dependencies of the form factors
are given as
\begin{eqnarray}
g(q^2) & = & \frac{1 \, GeV}{\left(1-\frac{q^2}{5.6^2}\right)^2} \, \, ,\, \,
f(q^2) = \frac{0.8 \, GeV}{\left(1-\frac{q^2}{6.5^2}\right)^2} \, \, , \, \,
g_1(q^2) = \frac{3. 74 \, GeV^2}{\left(1-\frac{q^2}{40.5}\right)^2} \, \, ,\, \,
f_1(q^2) = \frac{0.68 \, GeV^2}{\left(1-\frac{q^2}{30}\right)^2} ~.\nnb
\end{eqnarray}

We present the results of our analysis  in a series of figures.
Before their discussion  we give our SM predictions for the unpolarized BRs, for reference:
\bea
BR (B_s \rar \gamma \mu^+ \mu^-) & = & 1.52 \times 10^{-8}\, \, ,\nnb \\
BR (B_s \rar \gamma \tau^+ \tau^-) & = & 1.19 \times 10^{-8}\, \, ,\nnb
\eea
which are in  good agreement with the results of ref. \cite{Aliev3}.

\begin{figure}
\centering
\includegraphics[width=5in]{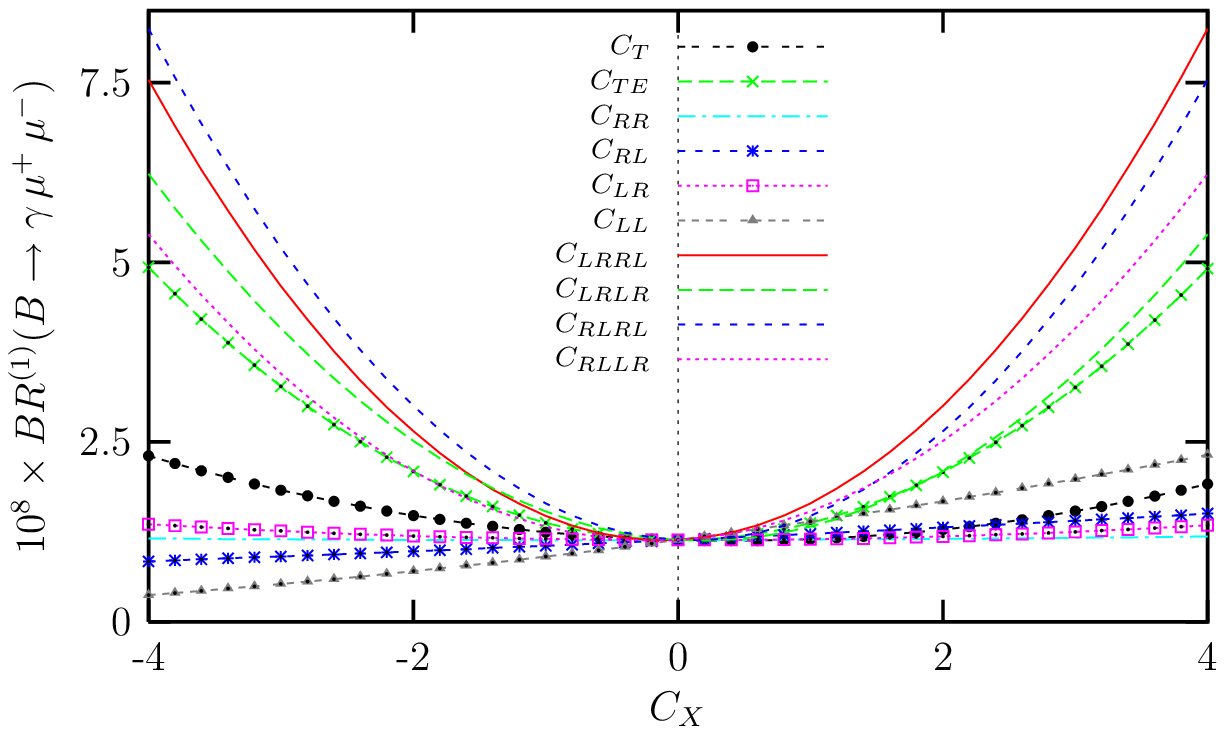}
\caption{The dependence of the integrated branching ratio for the
$B_s \rar \gamma \, \mu^+ \mu^-$  decay with photon  in the positive helicity state
on the new Wilson coefficients \label{f1}}
\end{figure}

\begin{figure}
\centering
\includegraphics[width=5in]{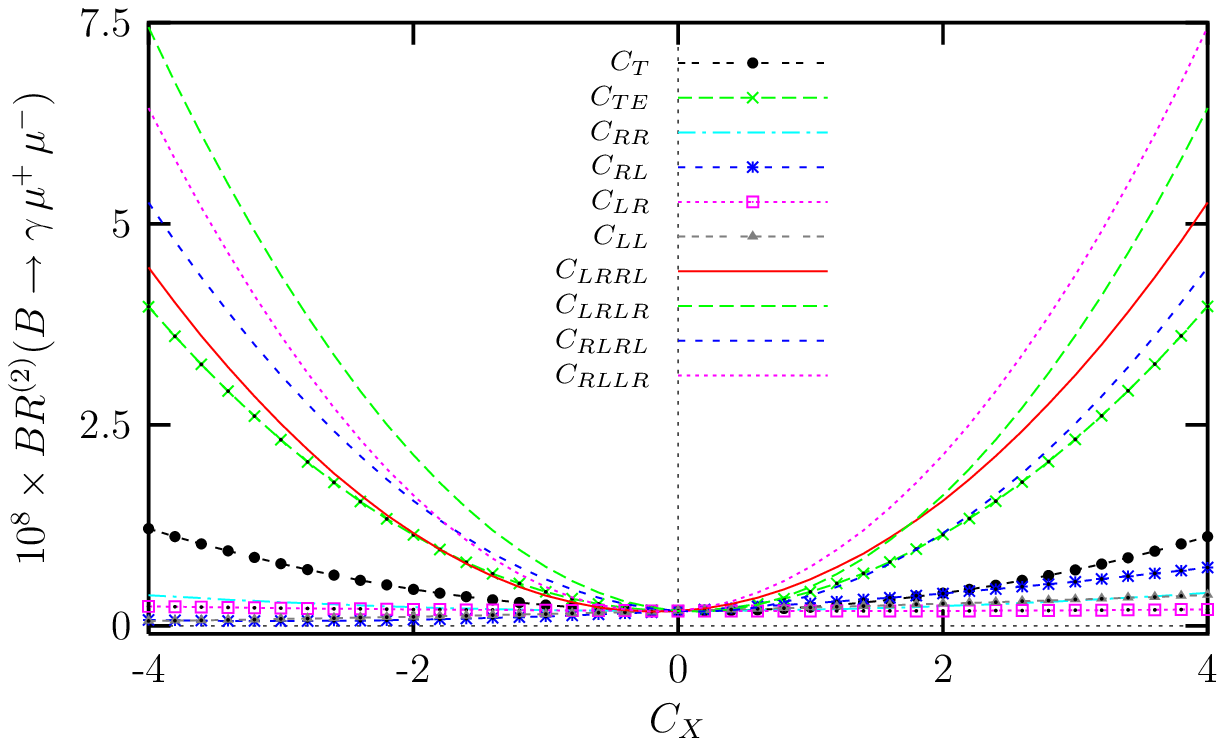}
\caption{The same as Fig.(\ref{f1}), but with photon in negative helicity state.\label{f2}}
\end{figure}

In Figs. (\ref{f1}) and (\ref{f2}), we present the dependence of the
$BR^{(1)}$ and $BR^{(2)}$  for $B_s \rar \gamma \mu^+ \mu^-$ decay
on the new Wilson coefficients, where the superscripts $(1)$ and $(2)$ correspond to the
positive and negative helicity states of photon, respectively. From these figures we see that
$BR^{(1)}$ and $BR^{(2)}$ are more sensitive to  all type of the scalar interactions
as compared to  the vector and tensor types; receiving the maximum contribution from the
one with coefficient $C_{RLRL}$ and $C_{LRLR}$, respectively. From Fig. (\ref{f2}),
we also observe that dependence of  $BR^{(2)}$ on all the new Wilson coefficients is symmetric
with respect to the zero point, while for $BR^{(1)}$, this symmetry is slightly lifted for the vector
type interactions (Fig.(\ref{f1})). It follows that  $BR^{(2)}$ decreases
in the region  $-4\le C_{X} \le 0$ and  tends to increase in between $0\le C_{X} \le +4$.
$BR^{(1)}$ exhibits a similar behavior, except for the vector  interactions with coefficients
$C_{LL}$, $C_{RL}$ and $C_{LR}$: it is almost insensitive to the existence of vector $C_{LR}$
type interactions and slightly increases with the increasing values of $C_{LL}$ and $C_{RL}$,
 receiving a value lower than the SM one between $-4$ and $0$.

\begin{figure}
\centering
\includegraphics[width=5in]{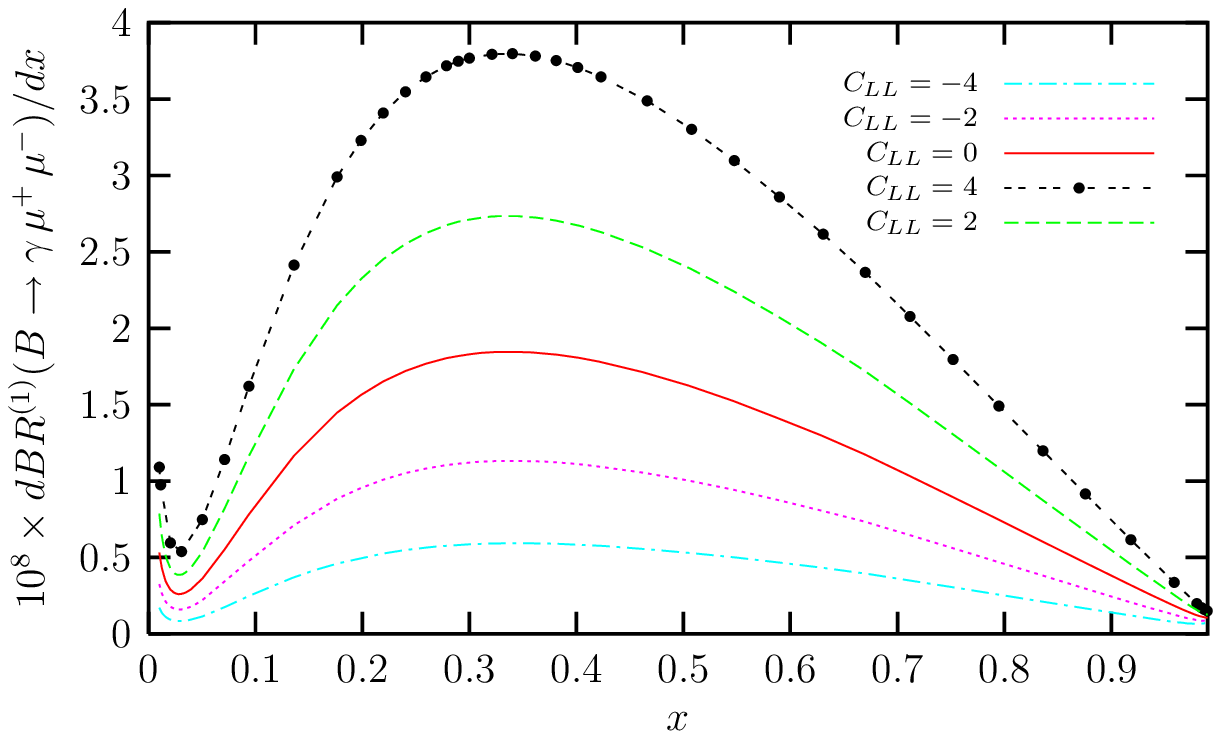}
\caption{The dependence of the differential branching ratio for the
$B_s \rar \gamma \, \mu^+ \mu^-$ decay with photon  in the positive helicity state
on the dimensionless variable $x=2 E_\gamma/m_B$ at different values of vector
interaction with coefficient $C_{LL}$ \label{f3}}
\end{figure}

\begin{figure}
\centering
\includegraphics[width=5in]{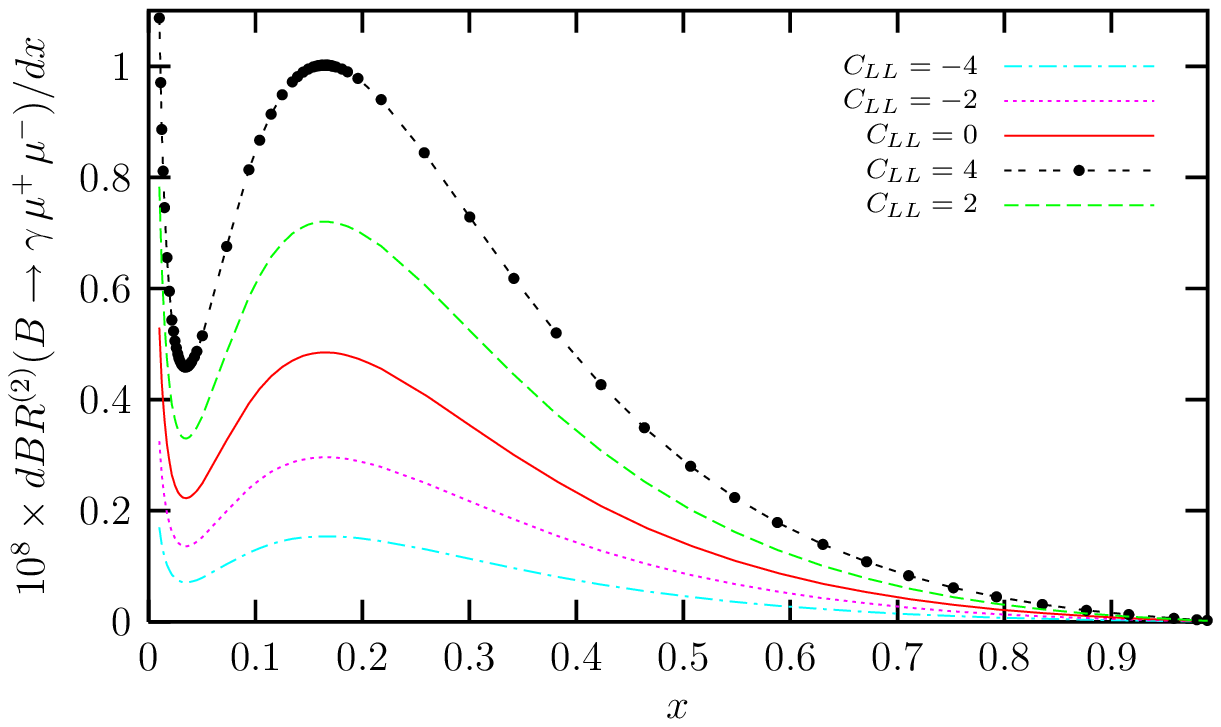}
\caption{The same as Fig.(\ref{f3}), but with photon in the negative helicity state.\label{f4}}
\end{figure}

\begin{figure}
\centering
\includegraphics[width=5in]{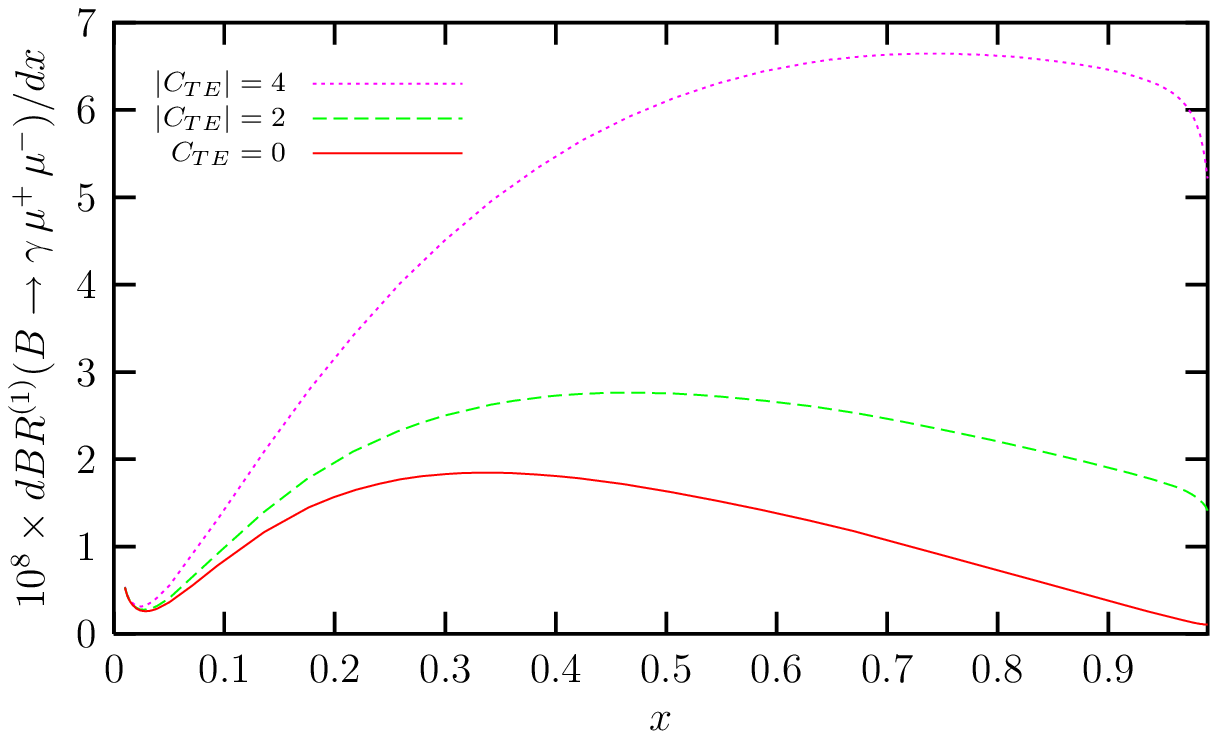}
\caption{The dependence of the differential branching ratio for the
$B_s \rar \gamma \, \mu^+ \mu^-$ decay with photon  in the positive helicity state
on the dimensionless variable $x=2 E_\gamma/m_B$ at different values of tensor interaction
with coefficient $C_{TE}$ \label{f5}}
\end{figure}

\begin{figure}
\centering
\includegraphics[width=5in]{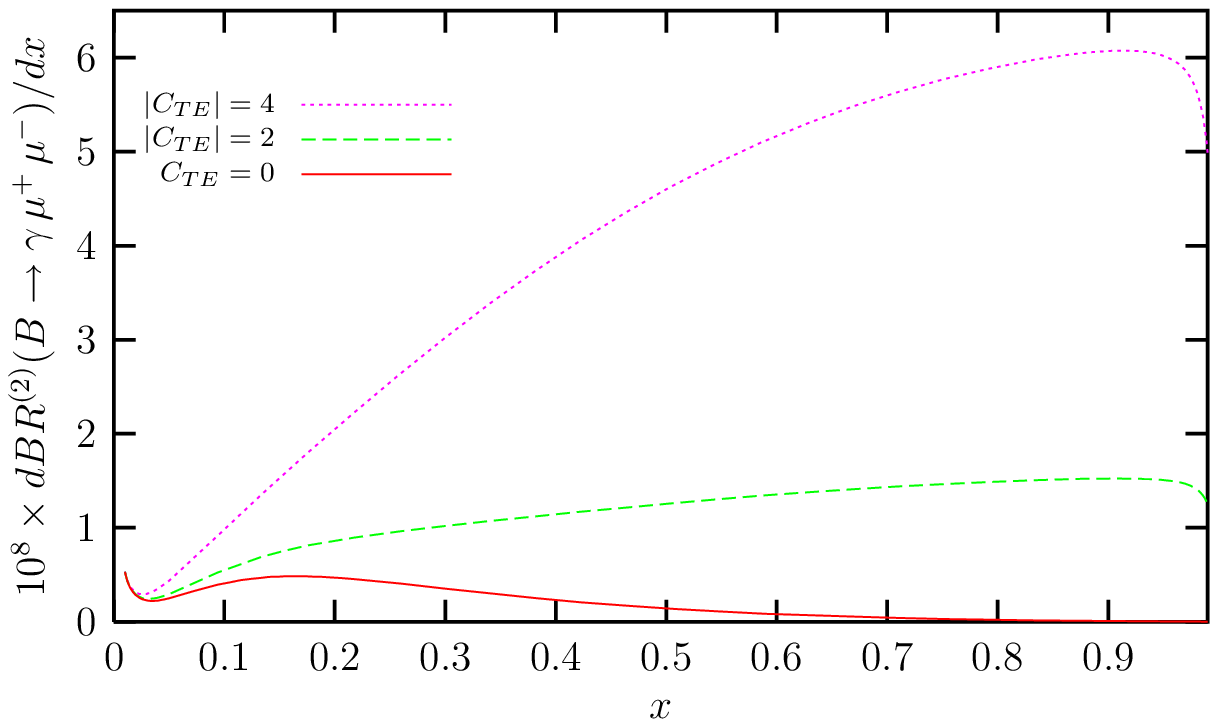}
\caption{The same as Fig.(\ref{f5}), but with photon in the negative helicity state.\label{f6}}
\end{figure}

\begin{figure}
\centering
\includegraphics[width=5in]{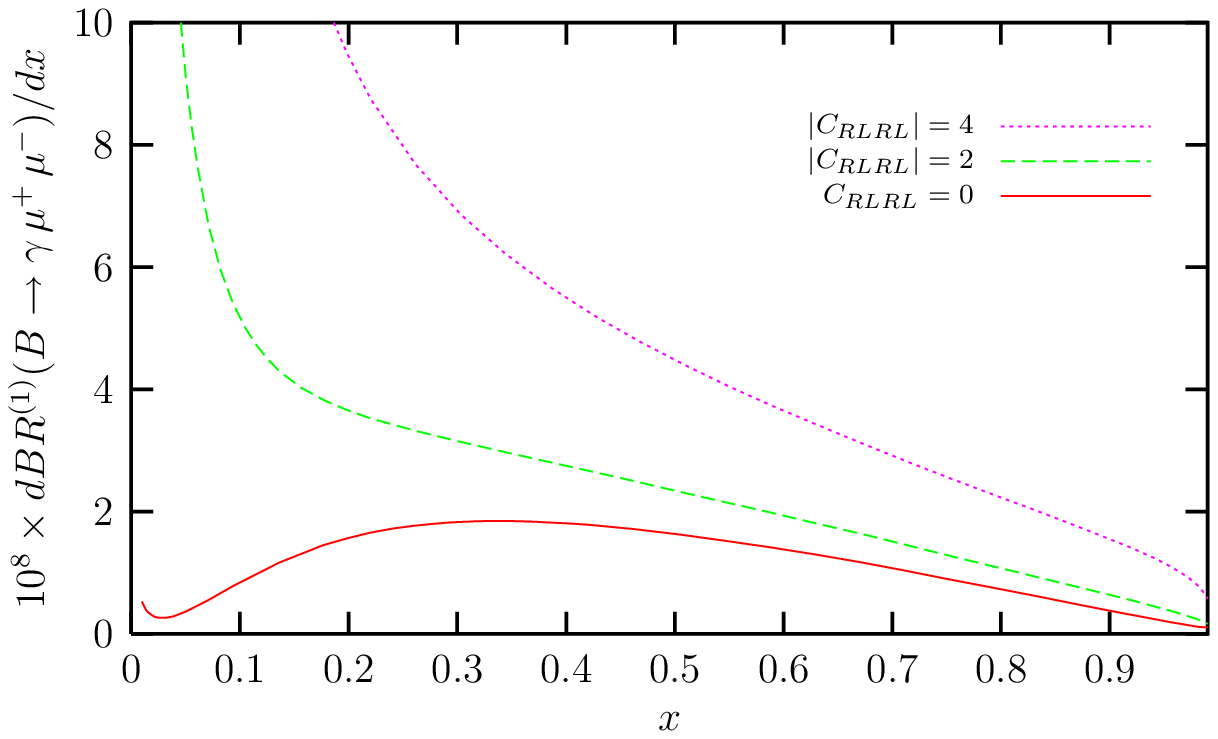}
\caption{The dependence of the differential branching ratio for the
$B_s \rar \gamma \, \mu^+ \mu^-$ decay with photon  in the positive helicity state
on the dimensionless variable $x=2 E_\gamma/m_B$ at different values of scalar interaction
with coefficient $C_{RLRL}$ \label{f7}}
\end{figure}

\begin{figure}
\centering
\includegraphics[width=5in]{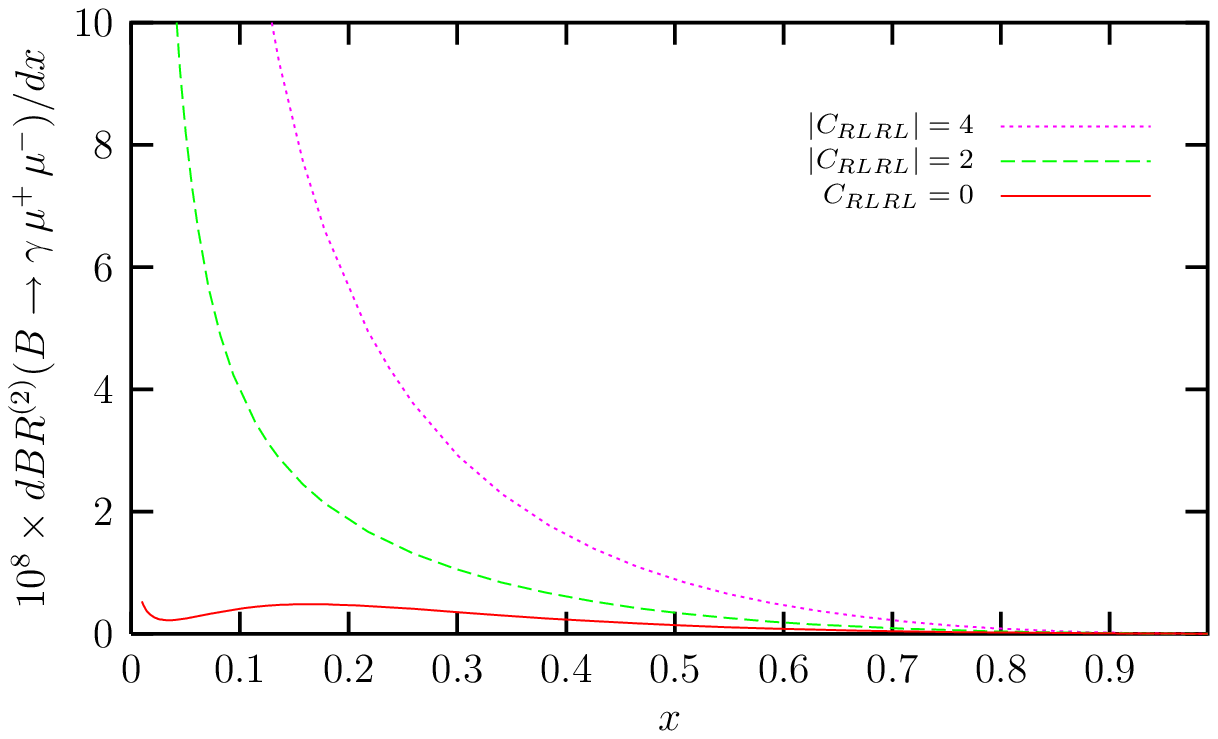}
\caption{The same as Fig.(\ref{f7}), but with photon in the negative helicity state.\label{f8}}
\end{figure}

Differential branching ratio can also give useful information about new
physics effects. Therefore, in Figs. (\ref{f3})- (\ref{f8}) we present the dependence of
the differential branching ratio with a polarized photon for the $B_s \rar \gamma \, \mu^+ \mu^- $ decay
on the dimensionless variable $x=2 E_\gamma/m_B$  at different values of vector, tensor and scalar
interactions with coefficients $C_{LL}$, $C_{TE}$ and $C_{RLRL}$. We observe  that
tensor (scalar) type interactions change the spectrum near the large (small)-recoil  limit,
$x \rar 1 $ $(x \rar 0)$, as seen from Figs.(\ref{f5},\ref{f6}) (Figs.(\ref{f7},\ref{f8})).
However, the vector type interactions increase the spectrum in the center of the
phase space and do not change it at the large or small-recoil  limit (Figs.(\ref{f3},\ref{f4})).
We also see from Figs.(\ref{f3}) and (\ref{f4}) that when $C_{LL} > 0$, the related
vector interaction gives constructive contribution to the SM result, but
for the negative values of  $C_{LL}$ the contribution is destructive. Therefore, it is possible to get the
information about the sign of new Wilson coefficients from measurement
of the differential branching ratio.

From Figs.(\ref{f1}-\ref{f8}), we also see that   the branching ratios
with a positive helicity photon  are greater than those with a negative helicity one.
To see this we rewrite Eq.(\ref{Deltapm}) for the SM in the limit $m_{\ell}\rar 0$,
\newpage
\bea
\Delta (\varepsilon _{i}) & = & \frac{m^2_B}{3} x^2 (-1+x)^2 \, \Bigg\{
 \Bigg |(C^{eff}_9-C_{10})(g \pm f) -\frac{2 C_{7}}{(1-x) m^2_B} m_b (g_1\pm f_1)\Bigg |^2
 \nnb \\
& + & \Bigg |(C^{eff}_9+C_{10})(g \pm f) -\frac{2 C_{7}}{(1-x) m^2_B} m_b (g_1\pm f_1)\Bigg |^2  \Bigg\}
\eea
where $+(-)$ is for $i=1(2)$. It obviously follows that $BR^{(1)} > BR^{(2)} $. We note that
this fact can be seen more clearly  from the comparison of the differential BRs for (1) and (2)
cases for the vector interactions with the coefficient $C_{LL}$, given in Figs.(\ref{f3}) and (\ref{f4}),
where $dBR^{(1)}/dx$ is larger about four times compared to $dBR^{(2)}/dx$.

\begin{figure}
\centering
\includegraphics[width=5in]{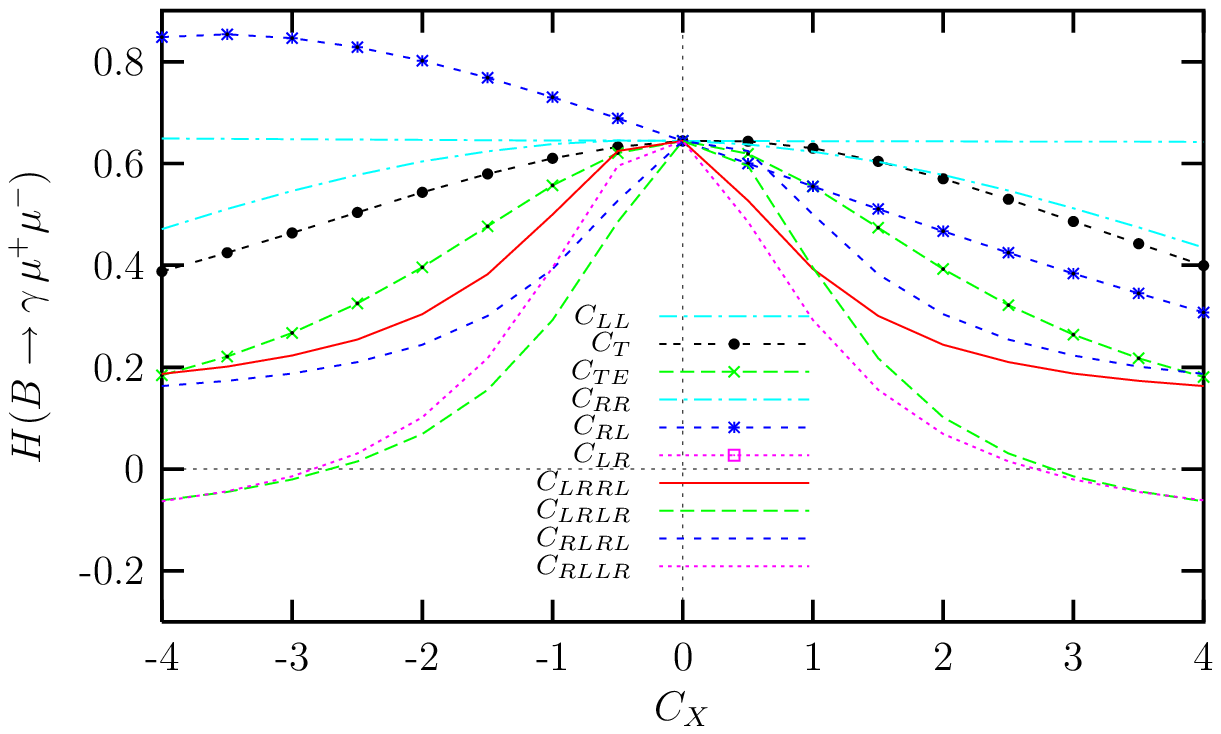}
\caption{The dependence of the integrated photon polarization asymmetry for
the $B_s \rar \gamma \, \mu^+ \mu^-$  decay on the new Wilson coefficients.\label{f9}}
\end{figure}

\begin{figure}
\centering
\includegraphics[width=5in]{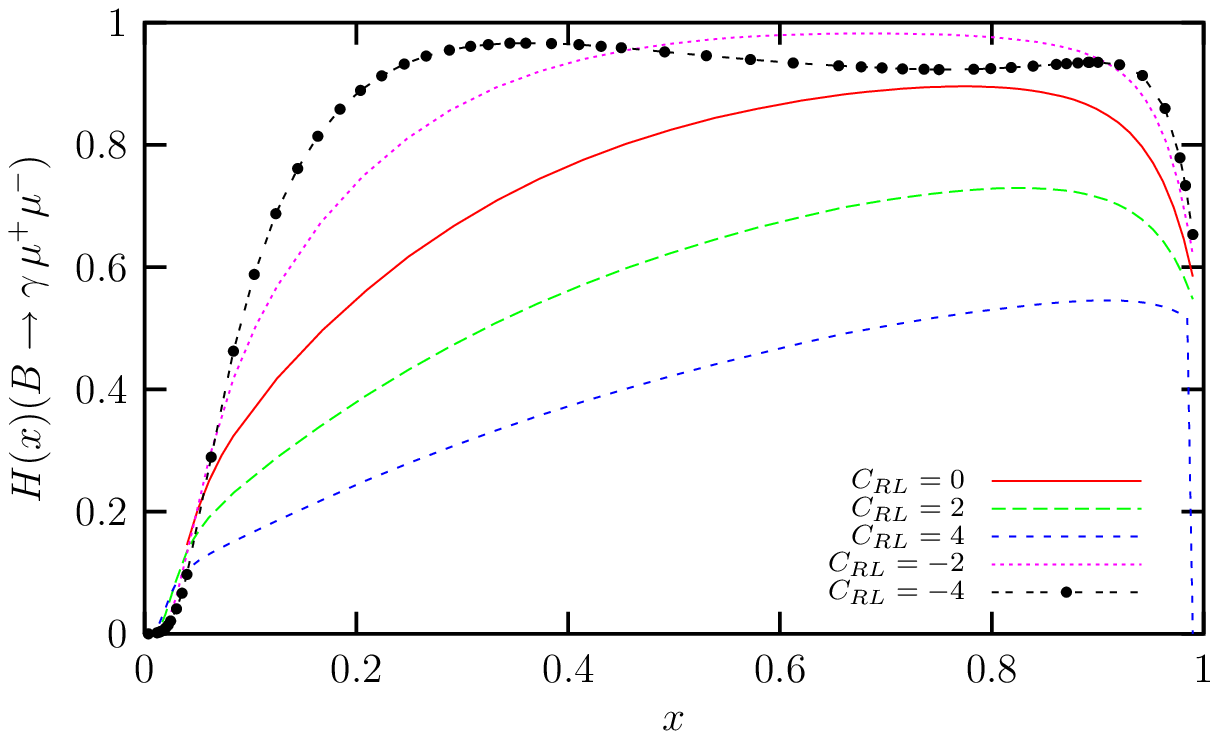}
\caption{The dependence of the differential  photon polarization asymmetry for
the $B_s \rar \gamma \, \mu^+ \mu^-$ decay on the dimensionless variable $x=2 E_\gamma/m_B$
for different values of $C_{RL}$.\label{f10}}
\end{figure}

In addition to the total and differential branching ratios, for
radiative decays like ours,  studying  the effects of
polarized photon may provide useful information about new Wilson coefficients.
For this purpose, we present the dependence of the integrated photon polarization
asymmetry $H$  for $B_s \rar \gamma \, \mu^+ \mu^- $ decay on the new Wilson coefficients
in Figs.(\ref{f9}) and  (\ref{f10}).
We see from Fig.(\ref{f9}) that spectrum of $H$
is almost symmetrical with respect to the zero point for all the new Wilson coefficients,
except the $C_{RL}$. The coefficient $C_{RL}$, when it is between $-2$ and $0$,
is also the only one  which gives the constructive contribution to the SM prediction
of $H$, which we find $H(B_s \rar \gamma \, \mu^+ \mu^- )=0.64$.
This behavior is  also seen  from  Fig.(\ref{f10}), in which
we plot the differential photon polarization asymmetry $H(x)$ for the same decay
as a function of $x$ for the different values of the vector  interaction with coefficients $C_{RL}$.
From these two  figures, we can conclude that performing measurement of $H$ at different photon energies
can give information about the signs of the new Wilson coefficients, as well
as their magnitudes.

\begin{figure}
\centering
\includegraphics[width=5in]{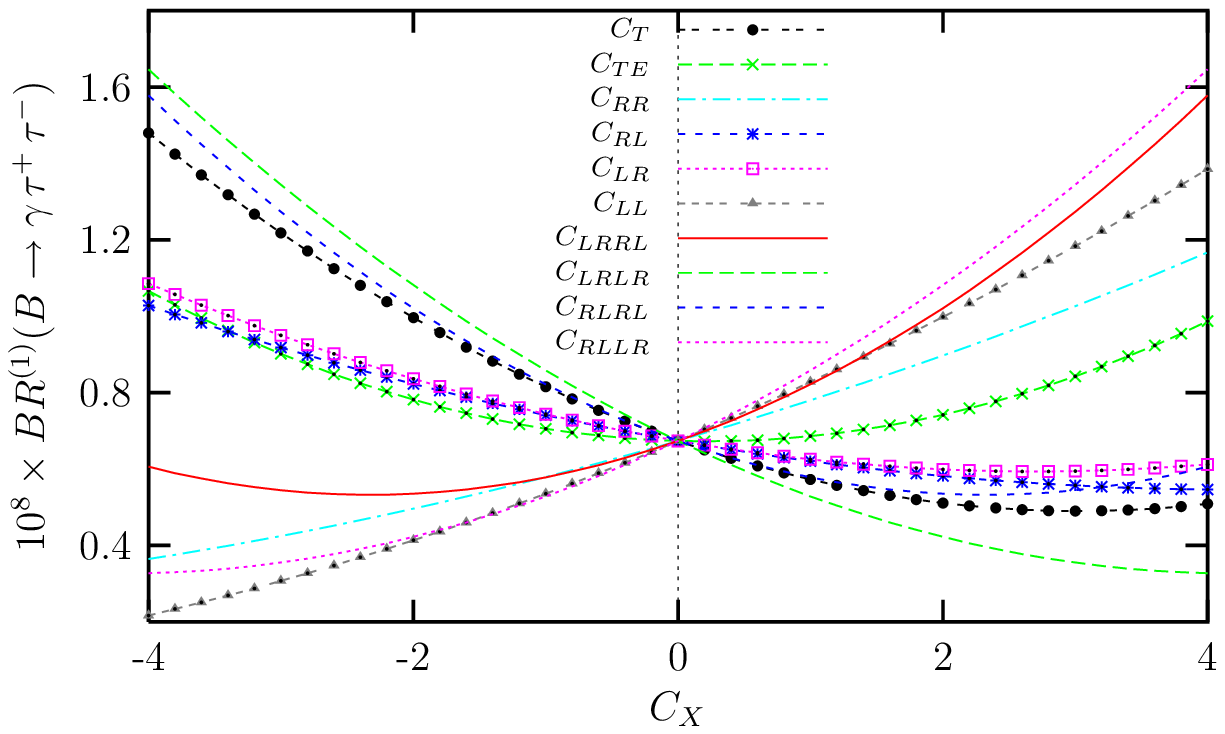}
\caption{The dependence of the integrated branching ratio for the
$B_s \rar \gamma \, \tau^+ \tau^-$  decay with photon  in the positive helicity state
on the new Wilson coefficients \label{f11}}
\end{figure}

\begin{figure}
\centering
\includegraphics[width=5in]{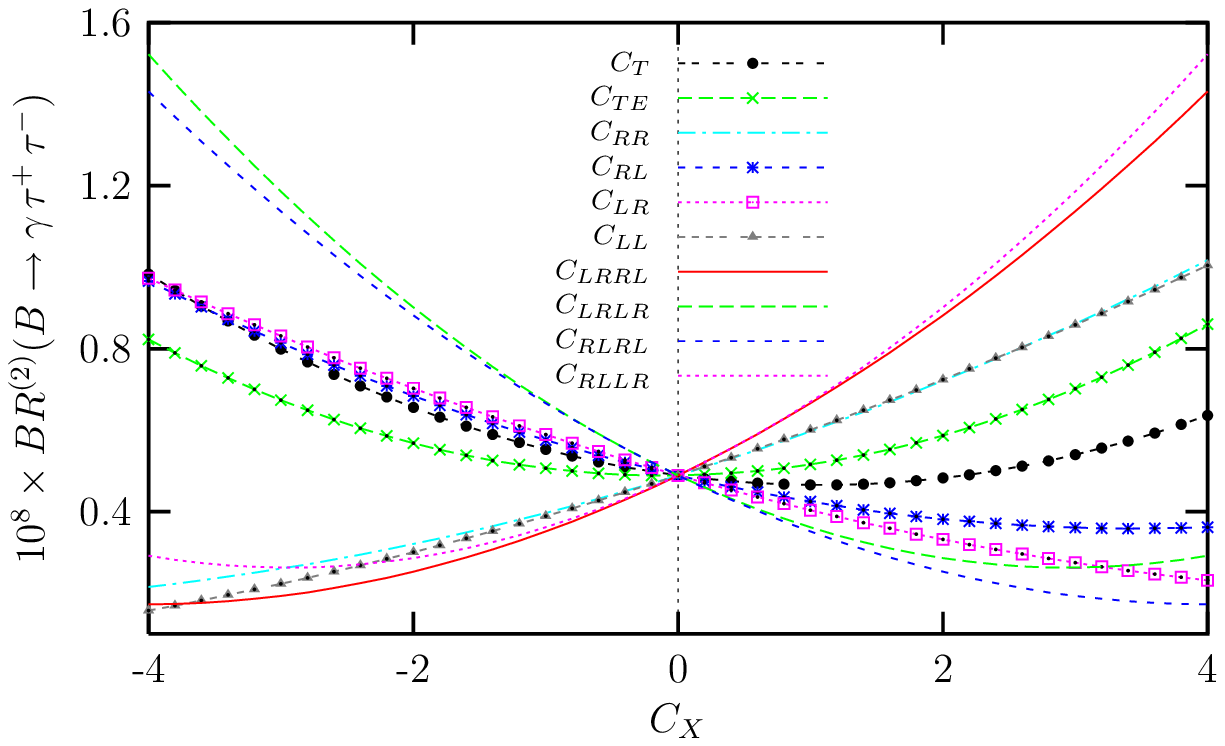}
\caption{The same as Fig.(\ref{f11}), but with photon in negative helicity state.\label{f12}}
\end{figure}

Note that the results presented in this work can easily be applied to the
$B_s \rar \gamma \,  \tau^+ \tau^- $ decay. For example,
in Figs.(\ref{f11}) and (\ref{f12}), we present the dependence of the
$BR^{(1)}$ and $BR^{(2)}$  for $B_s \rar \gamma \tau^+ \tau^-$ decay
on the new Wilson coefficients. We observe that in contrary to the $\mu^+ \mu^-$ final state,
spectrum of $BR^{(1)}$ and $BR^{(2)}$ for  $\tau^+ \tau^-$ final state is not symmetrical
 with respect to zero point, except for  the coefficient $C_{TE}$. Otherwise, we observe three
 types of behavior for $BR^{(2)}$ from Fig.(\ref{f12}): as the new Wilson coefficients $C_{LRRL}$, $C_{RLLR}$,
 $C_{LL}$ and $C_{RR}$ increase, $BR^{(2)}$ also increases. This behavior is reversed for
 coefficients $C_{LRLR}$, $C_{RLRL}$, $C_{LR}$ and $C_{RL}$, i.e., $BR^{(2)}$ decreases with
 the increasing values of these coefficients. However, situation is different for the tensor type
 interactions : $BR^{(2)}$ decreases when $C_{T}$ and $C_{TE}$
 increase   from $-4$ to $0$ and then increases in the positive half of the range.
We also observe from Fig.(\ref{f11}) that spectrum of $BR^{(1)}$ is identical to that of
$BR^{(2)}$ for the coefficients $C_{LRLR}$, $C_{LRRL}$, $C_{RLLR}$, $C_{LL}$, $C_{RR}$ and $C_{TE}$
in between $-4\le C_{X} \le +4$. For the rest of the coefficients, namely $C_{RLRL}$, $C_{LR}$, $C_{T}$,
it stand slightly below and almost parallel to the SM prediction in the positive half of the range,
although its behavior is  the same as $BR^{(2)}$ when $-4\le C_{X} \le 0$.

\begin{figure}
\centering
\includegraphics[width=5in]{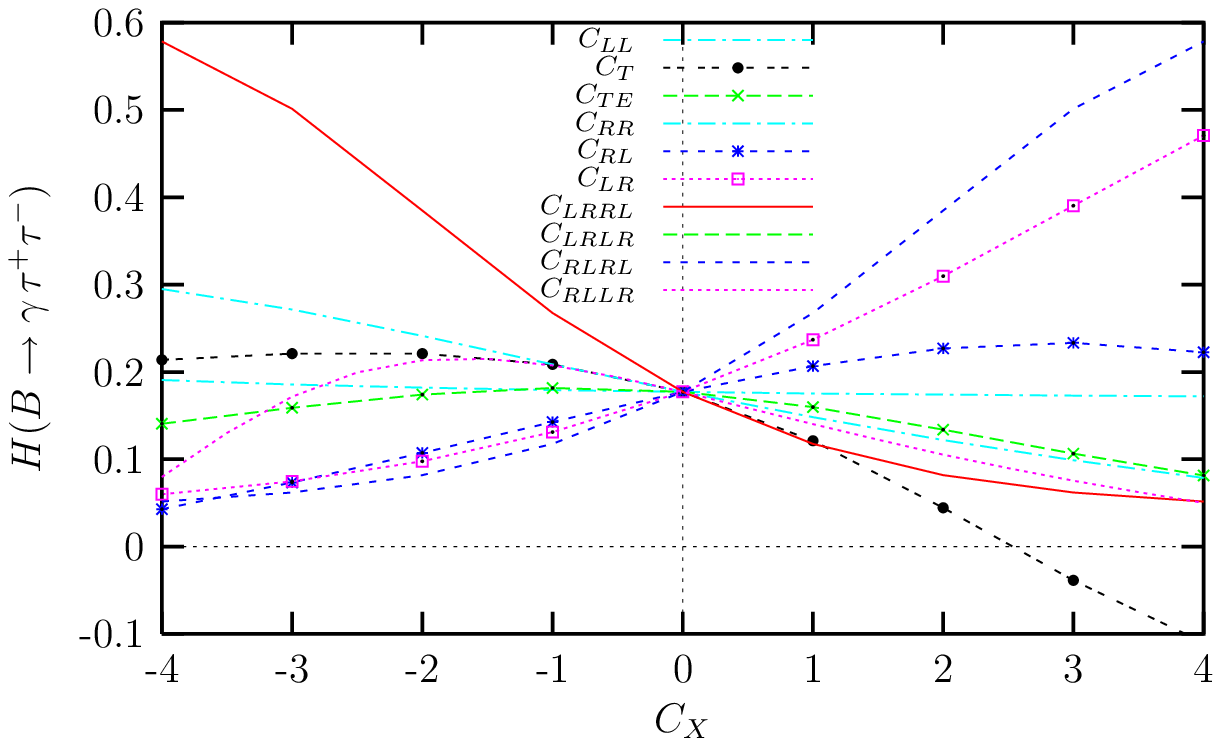}
\caption{The dependence of the integrated photon polarization asymmetry for
the $B_s \rar \gamma \, \tau^+ \tau^-$  decay on the new Wilson coefficients.\label{f13}}
\end{figure}

\begin{figure}
\centering
\includegraphics[width=5in]{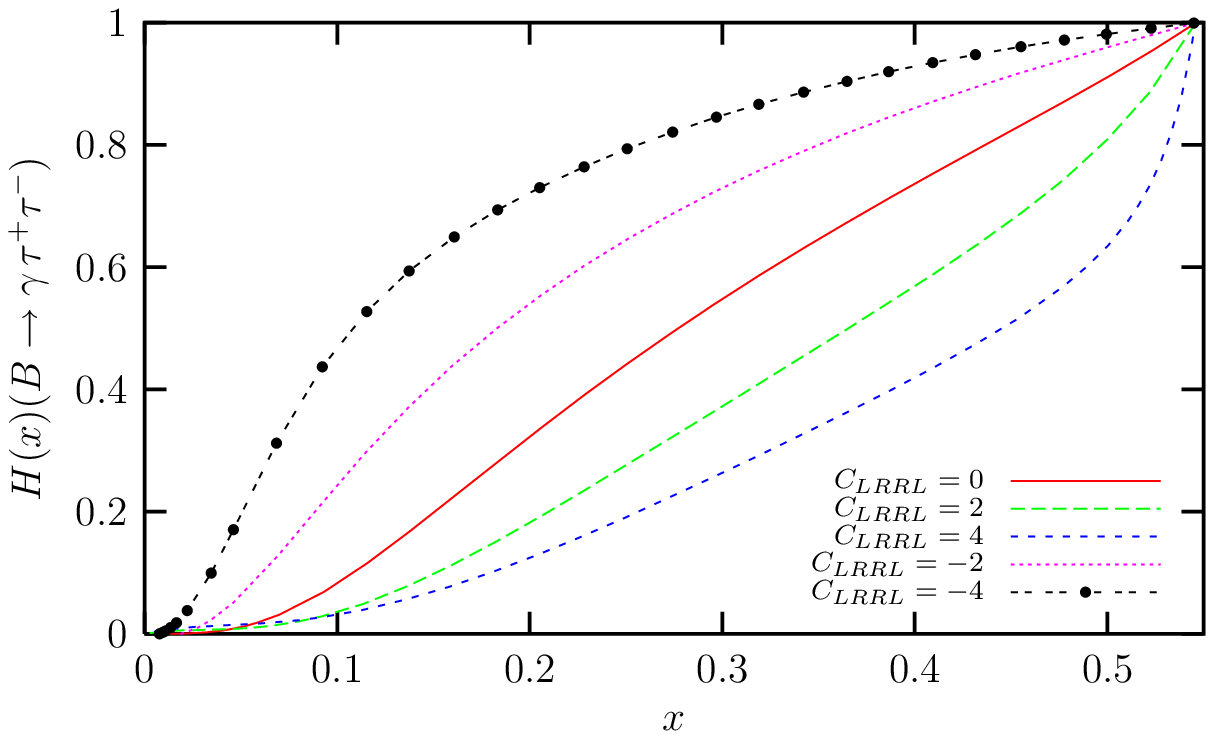}
\caption{The dependence of the differential  photon polarization asymmetry for
the $B_s \rar \gamma \, \tau^+ \tau^-$ decay on the dimensionless variable $x=2 E_\gamma/m_B$
for different values of $C_{LRRL}$.\label{f14}}
\end{figure}

Finally we present two more figures related to the   photon polarization
asymmetry $H$  for $B_s \rar \gamma \, \tau^+ \tau^- $ decay. Fig. (\ref{f13})
shows the dependence of the integrated photon polarization
asymmetry $H$  on the new Wilson coefficients. We present
 the differential photon polarization asymmetry $H(x)$ for the same decay
as a function of $x$ for the different values of the scaler  interactions with coefficients $C_{LRRL}$
in (\ref{f14}). We see from Fig. (\ref{f13}) that in contrary to the $\mu^+ \mu^-$ final state,
spectrum of $H$  for  $\tau^+ \tau^-$ final state is not symmetrical with respect to zero point.
It also follows  that when $0\le C_{X} \le 4 $ the dominant contribution to $H$ for
$B_s \rar \gamma \, \tau^+ \tau^- $ decay comes from $C_{RLRL}$ and $C_{LR}$. However,
for the negative part of the range $H$ receives constructive contributions mostly  from
$C_{LRRL}$, as clearly seen also from Fig.(\ref{f14}).

In conclusion, we have   studied the total and the differential branching ratios of
the rare $B_s \rar \gamma \,\ell^+ \ell^- $ decay  by taking into
account the polarization effects of the photon. Doing this we use a most general model independent
effective Hamiltonian, which contains the scalar and tensor type interactions as well as the vector types.
We have also investigated the sensitivity of "photon polarization asymmetry" in this radiative
decay to the new Wilson coefficients. It has been shown that all these physical observables  are very sensitive to the
existence of new physics beyond SM and their experimental measurements can give valuable information
about it.
\vspace{1cm}

\begin{center}
ACKNOWLEDGMENT
\end{center}
We would like to thank Prof. T. M. Aliev  for useful discussion.

\newpage
\renewcommand{\topfraction}{.99}
\renewcommand{\bottomfraction}{.99}
\renewcommand{\textfraction}{.01}
\renewcommand{\floatpagefraction}{.99}

\clearpage
\clearpage
\clearpage
\clearpage
\clearpage
\clearpage
\end{document}